# Adsorption of Metallic, Metalloidic, and Nonmetallic Adatoms on Two-Dimensional C$_3$N

*Meysam Makaremi,[1] Bohayra Mortazavi,[2] and Chandra Veer Singh\*[1,3]*

[1]Department of Materials Science and Engineering, University of Toronto, 184 College Street, Suite 140, Toronto, ON M5S 3E4, Canada.
[2]Institute of Structural Mechanics, Bauhaus-Universität Weimar, Marienstr. 15, D-99423 Weimar, Germany.
[3]Department of Mechanical and Industrial Engineering, University of Toronto, 5 King's College Road, Toronto M5S 3G8, Canada.

ABSTRACT: Two-dimensional polyaniline with a C$_3$N stoichiometry, is a newly fabricated material that has expected to possess fascinating electronic, thermal, mechanical and chemical properties . The possibility of further tuning the C$_3$N properties upon the adsorption of foreign adatoms is thus among the most attractive researches. We carried out extensive *ab-initio* density functional theory (DFT) simulations to investigate the adsorption of various elements including nonmetallic, metalloidic and metallic elements on the C$_3$N monolayer. While pristine C$_3$N acts as a semiconductor with an indirect electronic band gap; the functionalization with nonmetallic and semimetallic elements leads to a p-type doping and induces metallic behavior to the monolayer. On the other hand, metallic adsorption depending on the adatom size and the number of valence electrons may result in semiconducting, half-metallic or metallic properties. Whenever metallic foreign atoms conduct metallic characteristics, they mostly lead to the n-type doping by electron donation to the surface. Moreover, adsorption of transition metals could enhance the magnetic behavior of the monolayer due to the contribution of *d* electronic states. These results suggest that C$_3$N illustrates viable electronic-magnetic properties which could be promising for semiconducting, nanosensores and catalytic applications.

## 1. Introduction

A new class of materials called two-dimensional (2D) crystals has attracted extensive attention since the extraction and characterization of graphene from bulk graphite at controlled conditions in the last decade.[1] These single atomic layer crystals provide exceptional chemical, physical and mechanical characteristics quite distinctive from their original 3D bulk crystals.[2-5] During the past decade, most of 2D monolayer structures such as graphene, hexagonal boron-nitride (h-BN) and $MoS_2$ have been extracted from layered bulk materials composed of weekly Van der Waals bound stacked layers. Since the number of these layered materials is limited, the extracted 2D crystals are restricted to a handful of species. To resolve this issue, a new generation of these materials called synthetic 2D crystals has been recently grown and fabricated. The fabrication process does not require the initial bulk counterparts, and one can gain complementary functionalities and characteristics through designing and tuning the fabrication method.[6,7] In recent years, several honeycomb monolayers from group IV of elements; such as silicene, germanene, and stanene; have been synthesized and grown on different substrates.[8-12]

Several approaches have been developed to modify the electro-magnetic properties of 2D monolayers after the fabrication. These methods involve substitutional doping, defect engineering, functionalizing with adatoms, applying electric field and strain, among which surface functionalization can be considered as a robust approach.[13-18] Several computational studies have been conducted in order to investigate adatom adsorption on 2D monolayers. In one of the first studies in this field, Chan et al. investigated the adsorption of foreign elements on graphene by using DFT simulations.[19] The functionalization of silicene by metallic adatoms was studied and found that as a consequence of the buckled structure and the different surface reactivity, the interaction of silicene with adatoms is not comparable with that of graphene.[20] Adsorption properties of germanene, stanene, tin sulfide, and graphene derivatives interacting with different foreign atoms consisting of alkali, alkaline-earth, and transition metallic elements, and nonmetallic elements were the subjects of next investigations.[21-25]

Most recently, 2D polyaniline ($C_3N$) has been synthesized by the direct pyrolysis of hexaaminobenzene (HAB) trihydrochloride single crystals, and it has been characterized by scanning tunneling microscopy and scanning tunneling spectroscopy.[26] This novel material is predicted to offer good performance in a variety of applications; such as solution-membrane interfaces[27], solar cell devices[28], and electrolyte gating and doping of transistors[29], as a consequence of the stoichiometric formula of $C_3N$ and graphene-like structure in which nitrogen uniformly is distributed.[30-31] Recent theoretical studies confirmed that $C_3N$ can yield ultra-high stiffness and thermal conductivity.[32-33] As can be seen in Figure 1, the $C_3N$ honeycomb structure consists of two types of hexagonal rings including the CC rings with six carbon atoms and the NC rings with two nitrogen and four carbon atoms. The C-C and C-N bonds, both involve a similar atomic bond length of 1.40 Å. $C_3N$ involves a monoclinic unitcell with lattice constants a = b = 4.8614 Å containing 8 atoms (6 C and 2 N atoms). Functionalization of $C_3N$ monolayer by foreign atoms is expected to result in fascinating properties for this new 2D crystal, but has not been studied so far.

In the present work, we performed first principles electronic structure calculations to probe the role of various functionalizing elements consisting of metallic adatoms; such as alkali (AM), alkaline earth (AEM), Al and *3d* transition metallic (TM) elements, and semimetallic/nonmetallic adatoms; such as H, B, C, N, O, F, and P., on electronic-magnetic characteristics of $C_3N$. Properties including the binding energy, total density of electronic states (TDOS), partial density of electronic states (PDOS), Bader charge transfer, and the total magnetization were investigated in detail for each adatom. We first calculated the electronic structure properties of the bare monolayer. Next the adsorption of metalloids and nonmetals on the monolayer was considered, and finally we studied the electronic-magnetic properties of the $C_3N$ surface functionalized by various metallic species and compared with those of the former species.

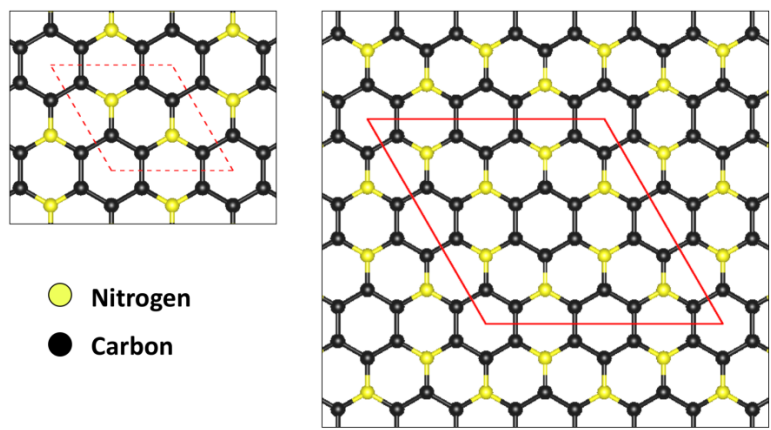

Figure 1. $C_3N$ unitcell (dashed red line), and 2x2 supercell (solid red line).

## 2. Computational Details

Using the Vienna *Ab-initio* Simulation Package (VASP)[34], we carried out first principle simulations based on spin polarized density functional theory (DFT) to study the interaction of the $C_3N$ monolayer with various adsorbing foreign atoms. Projector augmented-wave (PAW) potentials[35] and generalized gradient approximation (GGA) with the Perdew–Burke–Ernzerhof (PBE) functionals[36] were used to describe pseudopotential and exchange correlation functional terms, respectively. For Alkali atoms Na and K, we considered *p* semi-core states as valence states by choosing the Na_pv and K_pv pseudopotentials, respectively. The van der Waals DFT-D2[37] correction method of Grimme was applied to modify binding energy calculations. In addition to PBE simulations, Heyd-Scuseria-Ernzerhof (HSE06)[38] functionals were used to calculate the density of states for some cases. The Brillouin zone integration was performed through the Monkhorst-Pack scheme[39] with a mesh grid of 15x15x1. A kinetic energy cutoff of 500 eV and an electron self-consistent convergence criterion of 1x10$^{-6}$ eV were considered for calculations. For ionic relaxation the conjugate gradient method is applied with the force convergence criterion of 1x10$^{-3}$ eV/Å to minimize the Hellmann–Feynman forces. The tetrahedron method with Blöchl corrections was used as the smearing scheme.

As shown in Figure 1, a 2x2 supercell including 24 carbon and 8 nitrogen atoms was used for all simulations, and a vacuum space of 20 Å was applied in the sheet normal direction. To find the optimal size for the vacuum space, the convergence test with respect to the box dimension in the z direction was performed (see Figure S1), and the effect of adatom adsorption on the work function was probed for three different elements including P, Li, and Na by calculating the Fermi level energy and the average planar energy in the z direction (see Figure 2). The grid-based Bader scheme[40] was used to analyze the charge distribution, and charge difference resulting from the functionalization is explained as,

$$\Delta \rho_{Ads} = \rho_{C_3N+Atom} - \rho_{C_3N} - \rho_{Atom}, \qquad (1)$$

here, $\rho_{C_3N+Atom}$, $\rho_{C_3N}$, and $\rho_{Atom}$ are the total charges of the C$_3$N/adatom system, the pristine monolayer and the isolated adatom, respectively. In addition, adsorption energy of the adatom on the C$_3$N surface is calculated by,

$$E_{Ads} = E_{C_3N+Atom} - E_{C_3N} - E_{Atom}, \qquad (2)$$

where $E_{C_3N+Atom}$, $E_{C_3N}$, and $E_{Atom}$ are the total energies of the C$_3$N/adatom system, pristine C$_3$N and the isolated adatom, respectively. A large negative value of E$_{Ads}$ signifies strong binding.

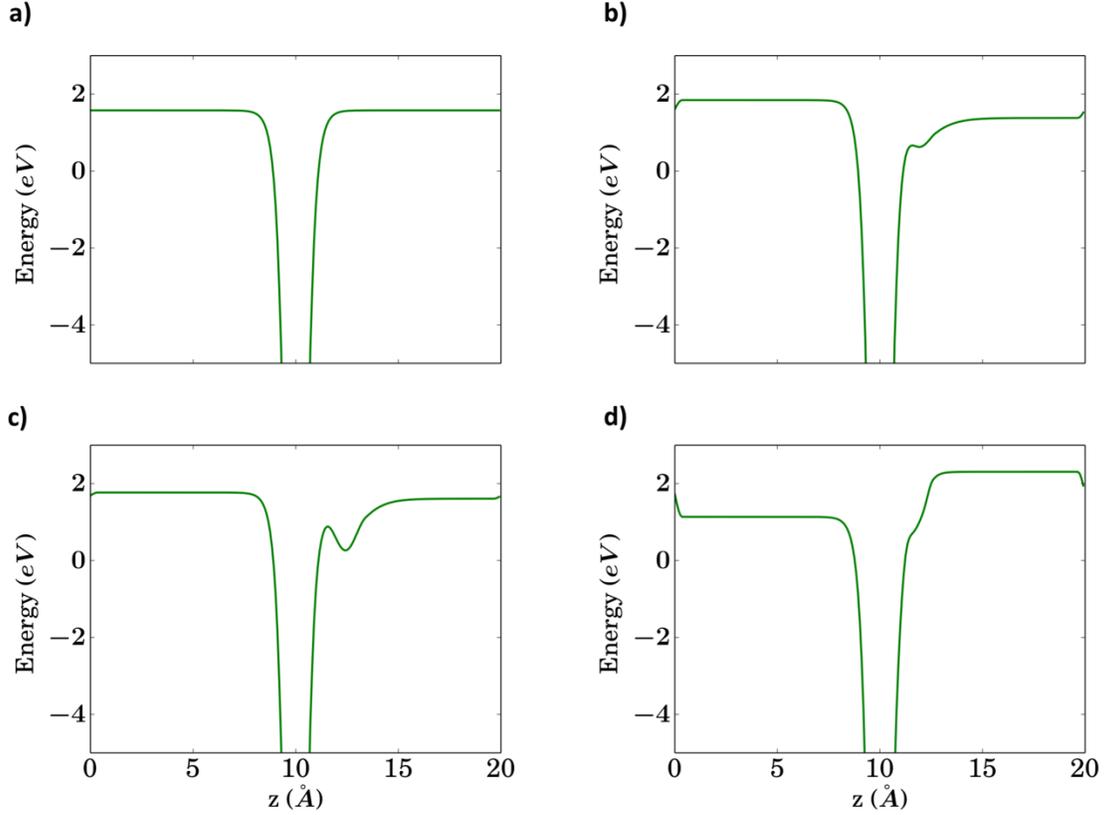

Figure 2. Average planar potential in the z direction for pristine $C_3N$ (a) and the $C_3N$ monolayer interacting with Li (b), Na (c), and O (d). The work function ($\Phi$) can be calculated from $\Phi = E_{vac} - \epsilon_F$, in which $E_{vac}$ and $\epsilon_F$ are the vacuum-potential and Fermi-level energy, respectively. From this formula, we predeicted the work-function the bare monolayer, and functionalized monolayers with Li, Na, and O to be 3.11, 2.73, 2.76, and 3.21 respectively.

To study the effect of adatom adsorption on the magnetic characteristics of the monolayer, the magnetic ground state need to be identified. Since this property is quite sensitive to the initial conditions, we manually searched the ground state for each system by considering various magnetic configurations including different initial magnetic moments and net magnetizations, and later we examined the consistency of the results after relaxation. Otherwise, the structure involving the minimum energy was chosen as the most possible adsorption configuration.

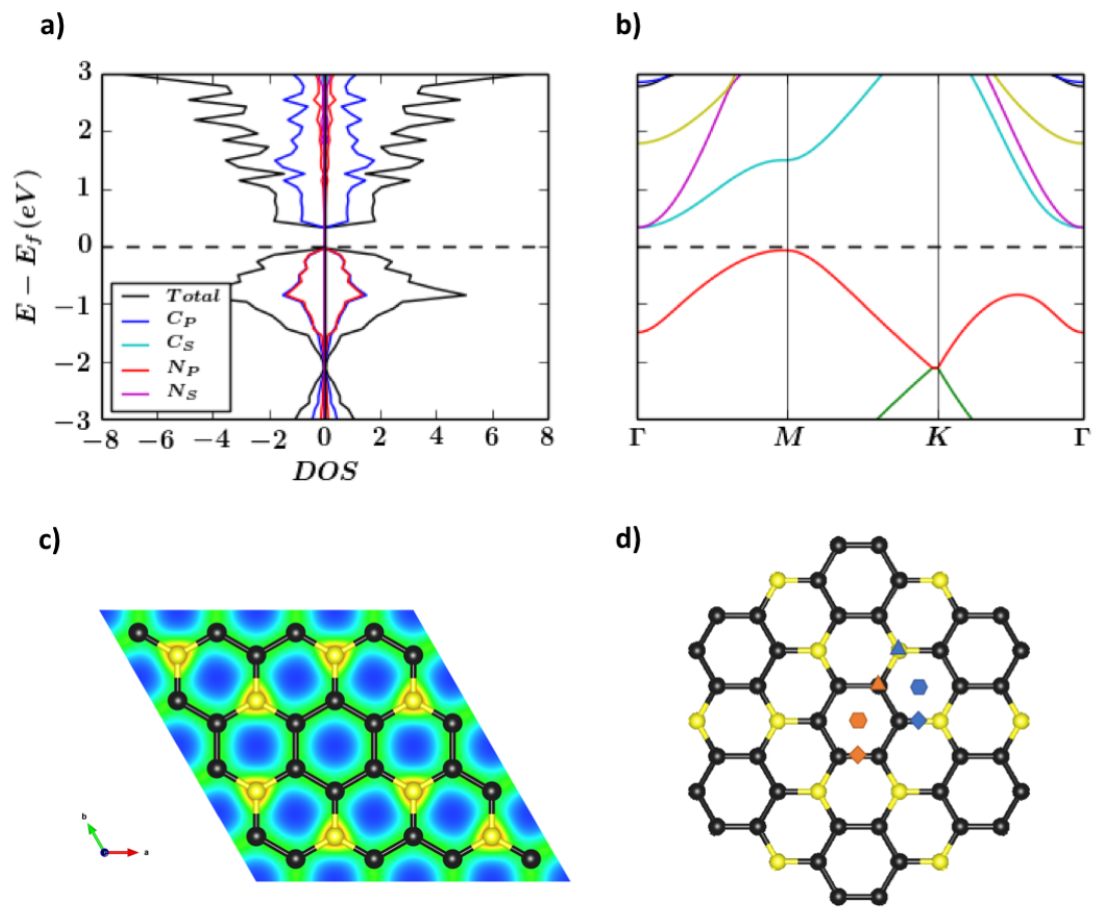

Figure 3. Electronic structure properties of $C_3N$. a) Spin polarized projected density of states. The black dashed line shows the fermi energy level. b) Electronic band structure. c) Electronic Charge distribution. d) Considered adsorption sites on the $C_3N$ surface, illustrated by orange, and blue diamonds for C-C and N-C bridge sites ($B_{CC}$ and $B_{NC}$); orange and blue hexagons for C-C and C-N hexagonal sites ($H_{CC}$ and $H_{NC}$); and orange and blue triangles for C and N tetrahedral sites ($T_C$ and $T_N$), respectively.

## 3. Results and Discussion

To analyze the functionalization process, it is always useful to probe the electronics properties of pristine $C_3N$ at the first step. Total density of states and projected density of states of the bare monolayer, illustrated in Fig. 3a, reveals that the valence band maximum is composed of *p* orbitals of both carbon and nitrogen, whereas the conduction band minimum is mainly derived by the contribution of the *p* orbitals of carbon atoms. Fig. 3b depicts the electronic band structure of $C_3N$, which suggests an indirect energy gap of $E_g = 0.39$ eV between the $\Gamma$ point and the M point resulting in semiconducting nature of the bare monolayer. It is well-known that PBE underestimates the energy gap, therefore we performed HSE06 functional calculations suggesting that the electronic density is similar to that of PBE, and only the conduction band is shifted upward, and leading to an energy gap of $E_g = 1.05$ eV (see Figure 4). The result is consistent with the literature[34,41] and verifies the accuracy of our modeling.

To study atomic adsorption, we investigated the adsorption effect of twenty-four different adatoms from different atomic species. As illustrated in Fig. 3d, six potential adsorption sites on the $C_3N$ surface were considered; 1) the bridge site above the midpoint of the C-C bond ($B_{CC}$), 2) the bridge site above the midpoint of the N-C bond ($B_{NC}$), 3) the hexagonal site on the center of the ring with six C atoms ($H_{CC}$), 4) the hexagonal site on the center of the ring composed of both C and N atoms ($H_{NC}$), 5) the tetrahedral site on top of the C atom ($T_C$), and the tetrahedral site on top of the N atom ($T_N$). To further ensure that the functionalized $C_3N$ nanomembranes are thermodynamically stable, after the relaxation step, we carried out DFT molecular dynamics simulations at 1000 °K for some systems involving P, Al, C, N, Mg, Cr, Co, Ni adatoms interacting with $C_3N$ $Site_{Ads}$, $E_{Ads}$, $D_{Ads}$, $E_g$, $\Delta\rho$, and $\mu$ show the adsorption site, the binding energy, the bond length between the adatom and the closest surface atom, the band gap, the Bader charge transfer from the adatom to $C_3N$, and the total magnetic moment, respectively. *Graphene data calculated by Ref. 42.

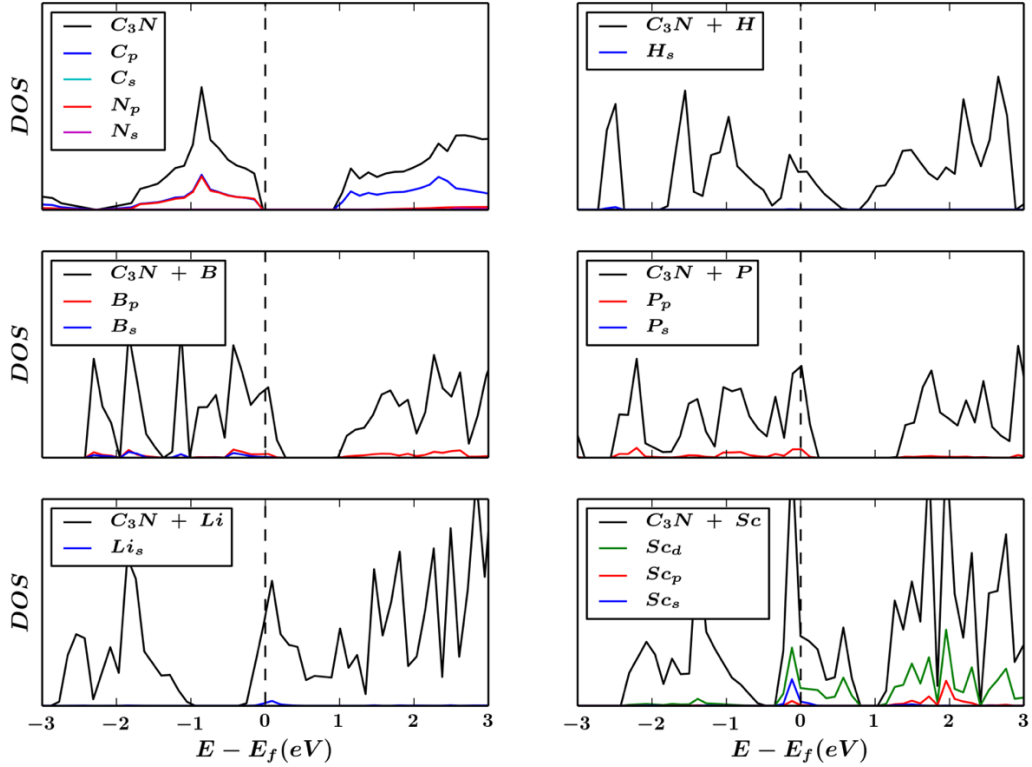

Figure 4. HSE06 density of states for the pristine monolayer and the monolayer functionalized by H, B, P, Li, and Sc adatoms.

Table 1: Electronic-magnetic properties of 2D $C_3N$ interacting with non-/semi-metallic adatoms.

| | $C_3N$ | | | | | | Graphene* | | |
|---|---|---|---|---|---|---|---|---|---|
| Element | $Site_{Ads}$ | $E_{Ads}$ [eV] | $D_{Ads}$ [Å] | $E_g$ [eV] | $\Delta\rho$ [e] | $\mu$ [$\mu_\beta$] | $Site_{Ads}$ | $E_{Ads}$ [eV] | $\Delta\rho$ [e] |
| H | $T_C$ | -1.64 | 1.12 | M | −0.04 | 0 | $T_C$ | -1.96 | −0.15 |
| B | $B_{CC}$ | -1.57 | 1.64 | M | −1.59 | 0 | $B_{CC}$ | -1.77 | −0.43 |
| C | $B_{CC}$ | -2.86 | 1.52 | M | +0.15 | 0 | $B_{CC}$ | -3.34 | +0.02 |
| N | $B_{CC}$ | -2.63 | 1.43 | M | +1.83 | 0 | $B_{CC}$ | -4.56 | +0.68 |
| O | $T_C$ | -3.64 | 1.30 | M | +1.79 | 0 | $B_{CC}$ | -4.79 | +0.84 |
| F | $T_C$ | -2.89 | 1.59 | M | +0.59 | 0 | $T_C$ | -2.90 | +0.59 |
| P | $B_{CC}$ | -1.26 | 1.88 | M | −1.87 | 0 | $B_{CC}$ | -2.20 | −0.38 |

**3.1. Adsorption of non/semimetallic foreign adatoms**

We investigated the role of common functionalizing nonmetallic and metalloidic agents including H, B, C, N, O, F and P on electronic-magnetic characteristics of $C_3N$. The results summarized in Table 1 indicate that the only carbon atoms of the monolayer act as binding sites, whereas no adsorption occurs on the nitrogen atoms. Among the studied nonmetallic and metalloidic agents, the most reactive one is O with a binding energy ($E_{Ads}$) of -3.64 eV as a consequence of its large number of valence electrons and short atomic radius, however the shortest adsorption bond distance ($D_{Ads}$) belongs to H which include the shortest atomic radius among the studied foreign atoms. It is worthwhile nothing that the species prefer to be adsorbed at $B_{CC}$ and $T_C$ sites of $C_3N$, similar to the way they interact with the graphene surface and also similar to graphene, $C_3N$ makes the strongest bonds with O, F, C and N.

Electronic density of states calculations of functionalizing nonmetallic and metalloidic species depicted in Figure 5 suggest that the interaction of the species with the monolayer eliminates the semiconducting band gap of pristine $C_3N$ and induces metallic properties. These adsorptions mostly produce a p-type carrier by shifting the Fermi level to the valence band edge. Listed in Table 1, the Bader analysis indicates that due to the electronegativity difference between the surface adsorbing atoms and the binding foreign atoms, the adsorption of nonmetallic and metalloidic atoms may result in a spectrum of charge transfer in which the maximal charge gain belongs to the phosphorus adsorption with a charge transfer of −1.87 e and the maximal charge loss happens due to the bonding of nitrogen including a charge induction of +1.83 e. (See Figure 6).

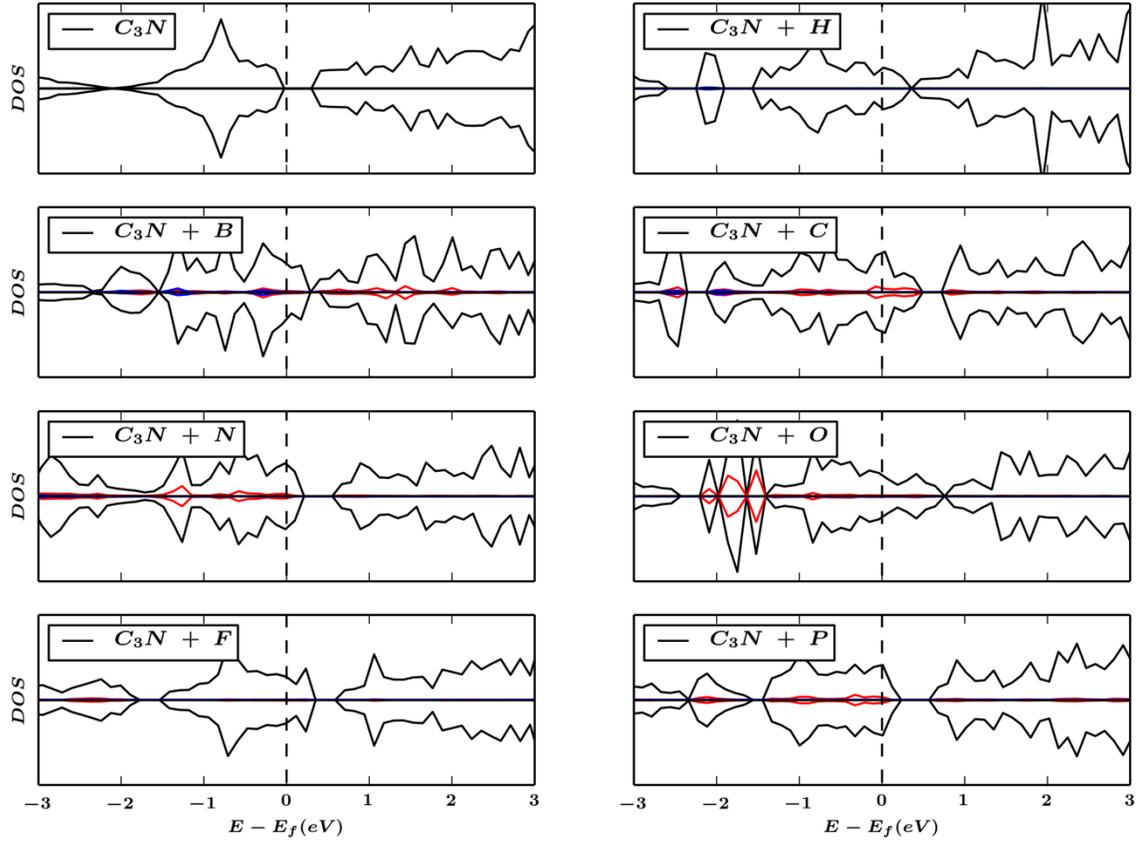

Figure 5. Density of states for the pristine and the functionalized $C_3N$ monolayer interacting with non-/semi-metallic elements. The black line shows TDOS, and Colorful lines illustrate the PDOS of functionalizing adatoms. Solid blue and red lines represent *s* and *p* states, respectively.

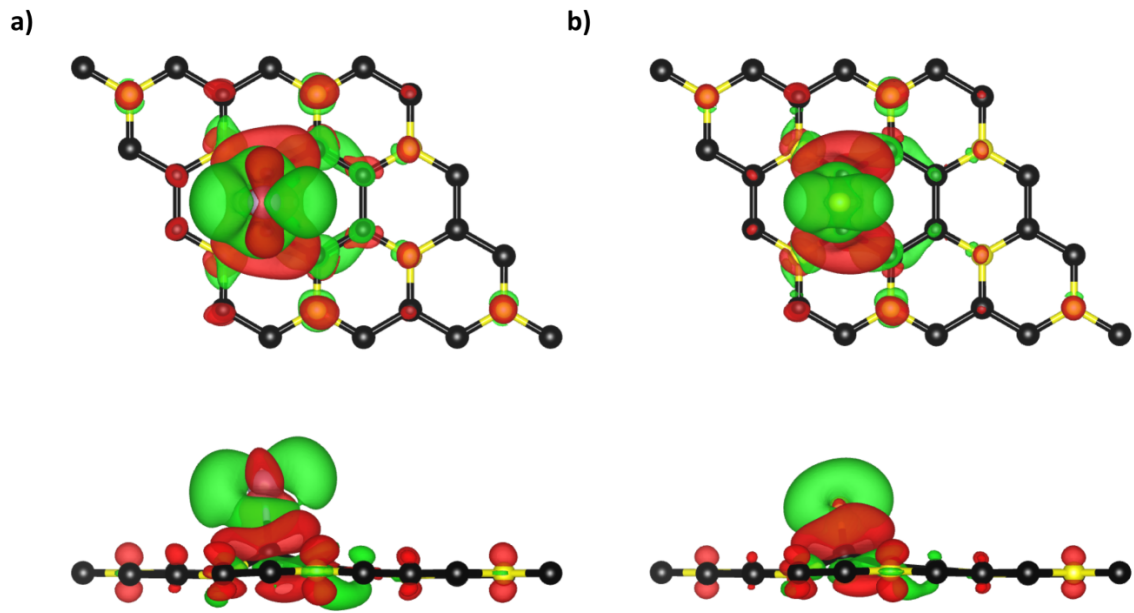

Figure 6. The differential charge density of the $C_3N$ monolayer functionalized by non-/semi-metallic adatoms: a) P, and b) N. Color coding consists of red for charge gain and green for charge loss.

## 3.2. Adsorption of metallic foreign adatoms

We calculated the adsorption properties of seventeen metallic elements including alkali (Li, Na and K), alkaline-earth (Be, Mg and Ca), group III (Al) and 3d transition (Sc, Ti, V, Cr, Mn, Fe, Co, Ni, Cu and Zn) metals. As illustrated in Table 2, similar to the former species, metallic ones avoid the nitrogen atoms of $C_3N$. Also metals tend to be adsorbed on $H_{CC}$ and $T_C$ sites while nonmetallic and metalloidic atoms prefer $B_{CC}$ and $T_C$ sites. (see Figure 7) which is similar to the interactions with graphene. The only exception occurs for Cu which prefers to be adsorbed on the $B_{CC}$ site of graphene. It is worthwhile to note that the adsorption on $B_{CC}$ and $T_C$ sites results in the local distortion of the monolayer where the adsorbing carbon atoms move toward the foreign atom.

Table 2: Electronic-magnetic properties of 2D $C_3N$ monolayer interacting with metallic adatoms.

| Element | C$_3$N | | | | | | Graphene* | | |
|---|---|---|---|---|---|---|---|---|---|
| | Site$_{Ads}$ | E$_{Ads}$ [eV] | D$_{Ads}$ [Å] | E$_g$ [eV] | $\Delta\rho$ [e] | $\mu$ [$\mu_\beta$] | Site$_{Ads}$ | E$_{Ads}$ [eV] | $\Delta\rho$ [e] |
| Li | H$_{CC}$ | -0.80 | 2.33 | M | –0.87 | 0.0 | H$_{CC}$ | -1.36 | –0.86 |
| Na | H$_{CC}$ | -0.33 | 2.77 | M | –0.71 | 0.0 | H$_{CC}$ | -0.72 | –0.62 |
| K | H$_{CC}$ | -0.44 | 3.09 | M | –0.63 | 0.0 | H$_{CC}$ | -0.81 | –0.63 |
| Be | H$_{CC}$ | -0.11 | 3.08 | 0.41 | –0.03 | 0.0 | H$_{CC}$ | -0.12 | –0.05 |
| Mg | H$_{CC}$ | -0.09 | 3.32 | 0.40 | –0.13 | 0.0 | H$_{CC}$ | -0.03 | –0.10 |
| Ca | H$_{CC}$ | -0.43 | 2.72 | M | –0.80 | 0.0 | H$_{CC}$ | -0.52 | –0.85 |
| Al | H$_{CC}$ | -0.89 | 2.57 | M | –1.25 | 0.0 | H$_{CC}$ | -1.62 | –0.81 |
| Sc | H$_{CC}$ | -1.84 | 2.30 | M | –1.24 | 1.1 | H$_{CC}$ | -2.08 | –1.10 |
| Ti | H$_{CC}$ | -2.25 | 2.30 | M | –1.05 | 2.8 | H$_{CC}$ | -3.27 | –1.10 |
| V | H$_{CC}$ | -1.36 | 2.13 | M | –1.13 | 1.0 | H$_{CC}$ | -3.88 | –0.98 |
| Cr | T$_C$ | -0.25 | 2.03 | 0.23 | –1.01 | 0.0 | H$_{CC}$ | -3.99 | –0.84 |
| Mn | H$_{CC}$ | -0.37 | 2.43 | HM | –0.48 | 5.0 | H$_{CC}$ | -3.82 | –0.70 |
| Fe | H$_{CC}$ | -1.39 | 2.09 | M | –0.61 | 2.0 | H$_{CC}$ | -3.83 | –0.58 |
| Co | H$_{CC}$ | -1.72 | 2.04 | M | –0.49 | 0.0 | H$_{CC}$ | -3.64 | –0.48 |
| Ni | H$_{CC}$ | -1.98 | 2.08 | 0.49 | –0.41 | 0.0 | H$_{CC}$ | -3.08 | –0.45 |
| Cu | T$_C$ | -0.71 | 2.03 | M | –0.08 | 0.1 | B$_{CC}$ | -0.97 | –0.19 |
| Zn | H$_{CC}$ | -0.24 | 3.30 | 0.38 | –0.04 | 0.0 | H$_{CC}$ | -0.13 | –0.03 |

Site$_{Ads}$, E$_{Ads}$, D$_{Ads}$, E$_g$, $\Delta\rho$, and $\mu$ show the adsorption site, the binding energy, the bond length between the adatom and the closest surface atom, the band gap, the Bader charge transfer from the adatom to C$_3$N, and the total magnetic moment, respectively. *Graphene data calculated by Ref. 42.

Since TM elements include the largest number of valence electrons and shortest atomic radii among studied metallic atoms, the ones that form the strongest bonds with the monolayer come from these elements including Ni, Ti, Co, Sc, Fe and V with binding energies of -1.89, -1.84, -1.70, -1.69, -1.39 and -1.38 eV, respectively. In addition, the foreign atoms from group AM, and group III make stronger interaction with C$_3$N compared to the ones from group AEM. The adsorption distance of metallic adatoms is larger than that of the other types as a consequence of their larger radii. Among the elements of the metallic type, Al and TM elements form shorter bonds to the surface due to their comparable small atomic radii.

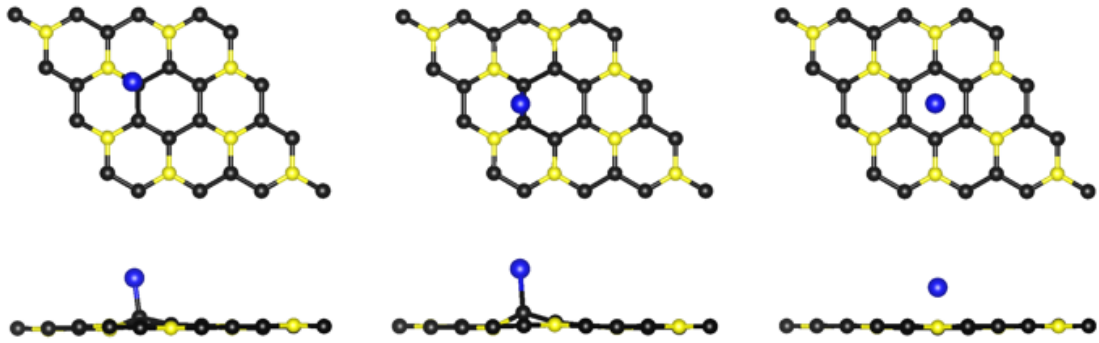

Figure 7. Adsorption configurations of the foreign adatom on C$_3$N. From left to right, atomic adsorption on T$_C$, B$_{CC}$, and H$_{CC}$ sites. The blue ball represents the adatom.

Table 2 suggests that functionalization by all Metallic species may not lead to metallic behavior of the monolayer. Among metallic adatoms, Be, and Mg from AE metals and Cr, Ni, and Zn from 3d T metals cannot change the semiconducting nature of C$_3$N and they involve energy gaps of 0.41, 0.40, 0.23, 0.49 and 0.38 eV, respectively. To further verify that functionalization can introduce conducting behavior to C$_3$N, we carried out HSE06 simulations for some systems including nonmetallic (H and P), semi-metallic (B) and metallic (Li and Sc) adatoms. As shown in Figure 4, our hybrid functional calculations illustrate that the considered systems have the zero energy gap which is in agreement with the PBE results.

The semiconducting and conducting behavior of functionalized monolayers suggest this 2D material as a potential candidate for many applications such as catalytic, solar cell and electronic devices. Interestingly adsorption of Mn from the TM group induces the half metallic properties composed of a conducting spin up channel and an insulating spin down channel with an electronic band gap of 0.43 eV which makes this functionalized monolayer potentially valuable for the spintronic applications. Metallic adsorption on C$_3$N mainly results in the charge acceptance of the monolayer (see Figure 8); while, in the nonmetallic and semimetallic functionalization, the monolayer both may donate and accept charges. This nature of $C_3N$ is consistent with the one from graphene as listed in Tables 1 and 2.

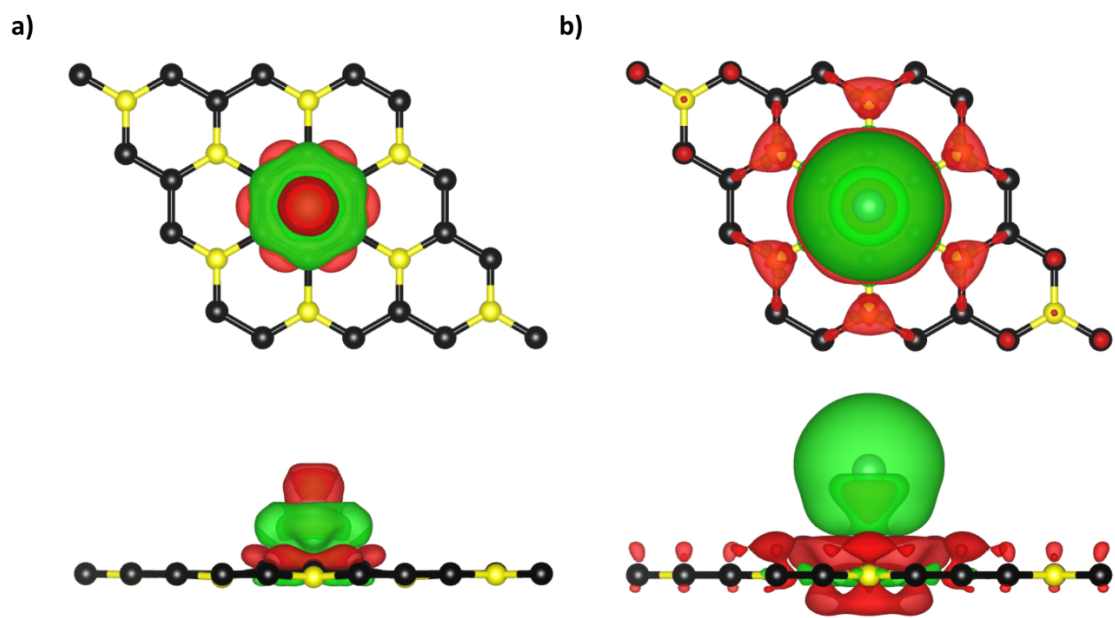

Figure 8. The differential charge density of the $C_3N$ monolayer functionalized by metallic adatoms: a) Ni and b) Be. Color coding of red and green illustrate the charge gain and charge loss, respectively.

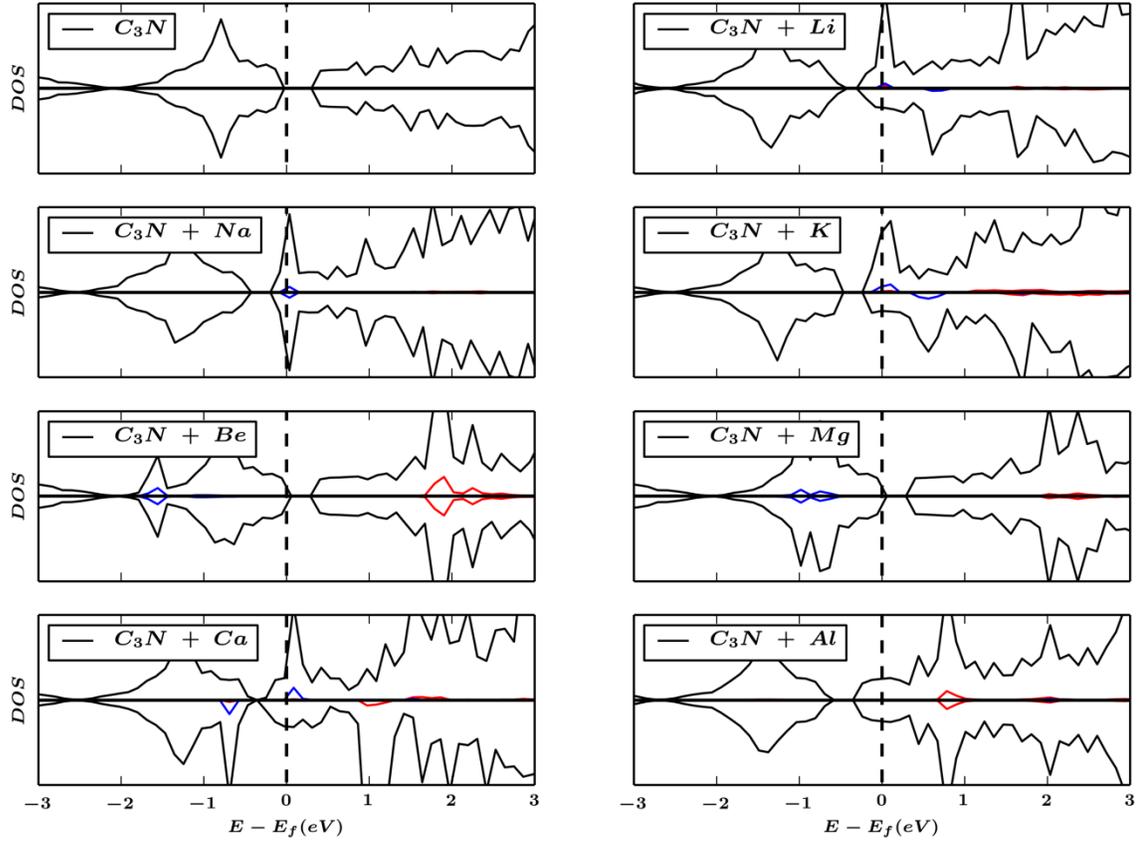

Figure 9. Density of states for the pristine and the functionalized $C_3N$ monolayer interacting with alkali, alkali earth and group III metallic elements. The black line shows TDOS, and Colorful lines illustrate the PDOS of functionalizing adatoms. Solid blue and red lines show *s* and *p* states, respectively.

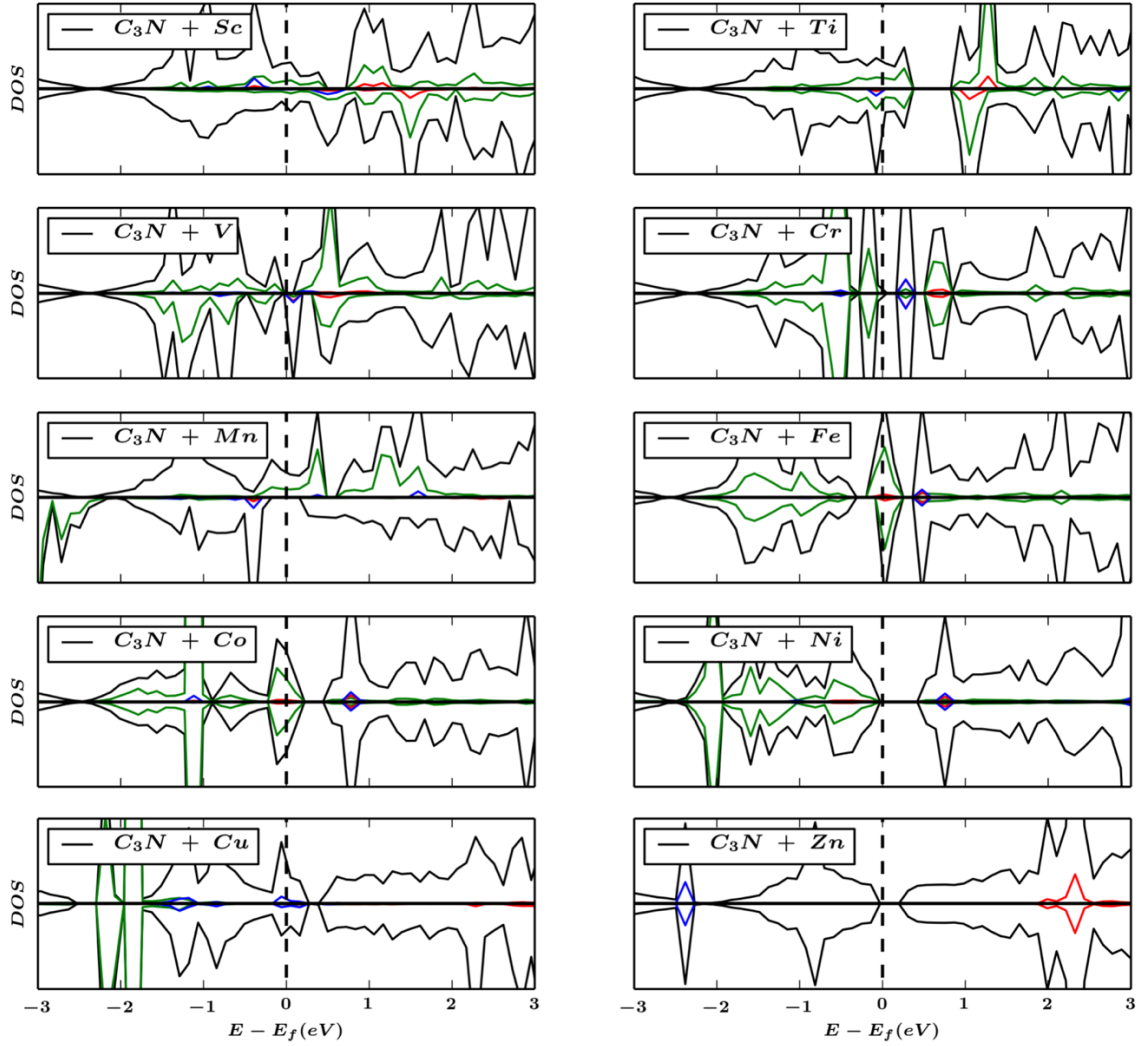

Figure 10. Density of states for the pristine and the functionalized $C_3N$ monolayer interacting with *3d* transition metallic elements. The black line represents TDOS, and Colorful lines illustrate the PDOS of functionalizing adatoms. Solid blue, red, and green lines shows $s, p$, and $d$ states, respectively.

Figures 9 and 10 illustrate the TDOS and PDOS profiles of $C_3N$ interacting with metallic foreign atoms. While the functionalization of nonmetallic and metalloidic adatoms leads to p-type doping, adsorption of many metallic ones from AM, EAM, and group III; such as Li, Na, K, Ca, and Al; mainly transfers the Fermi level to the conduction bond and causes n-type doping.

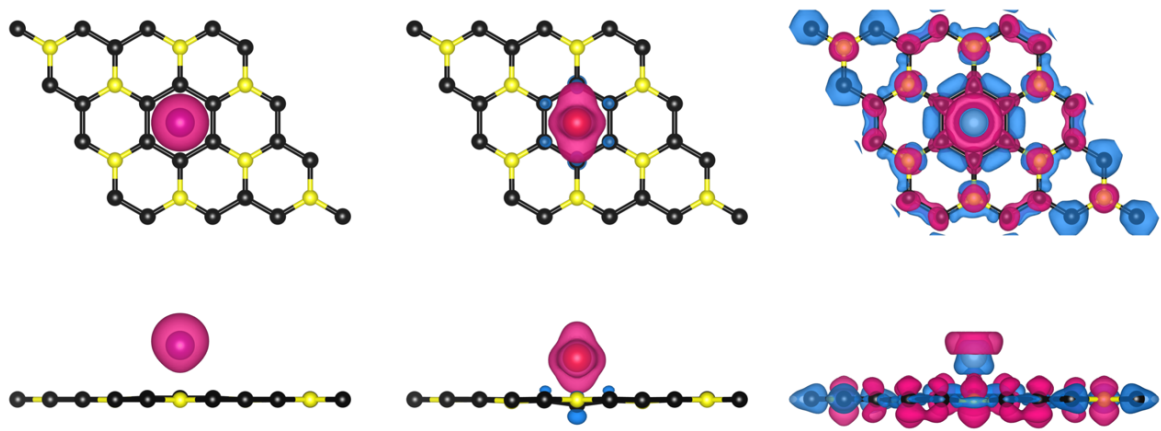

Figure 11. Different $C_3N$ magnetic configurations due to the adsorption of *3d* TM adatoms: Mn ($\mu = 5\ \mu_\beta$), V ($\mu = 1\ \mu_\beta$), and Zn ($\mu = 0\ \mu_\beta$) respectively (from left to right). Blue and pink represent opposite magnetization states.

As can be seen in Figure 10, *d* state electrons inducing magnetic properties play essential roles at the Fermi level during the adsorption of TM elements. Our magnetic moment calculations illustrate that the pristine $C_3N$ and the monolayers interacting with nonmetallic, metalloidic and metallic foreign atoms (except TM elements) present the nonmagnetic behavior. Whereas, the adsorption of most of the *3d* TM elements results in a nonzero magnetic moment (see Figure 11). The exceptions occur when $C_3N$ weakly interacts with Cr, or when the monolayer functionalized with Co, Ni, Cu and Zn in which the *3d* shell is almost filled, cancelling out the total magnetization and rendering symmetric spin-up and spin-down electronic density profiles. It should be noted that by doubling the supercell size in the x and y directions, one can study the nature of transferred magnetic properties *i.e.* ferromagnetic and antiferromagnetic to the $C_3N$ Monolayer.[43]

## 4. Conclusions

We used first principle PBE and HSE06 DFT calculations to investigate electronic-magnetic properties of a recently experimentally grown single layer 2D material ($C_3N$) functionalized by various adatoms. The calculations indicate that foreign adatoms prefer to be adsorbed on carbon sites of the monolayer. While, the adsorption of metallic elements mostly occurs at $H_{CC}$ and $T_C$ sites, other atomic species would be attracted to $B_{CC}$ and $T_C$ ones. The $C_3N$ surface accepts charge due to the adsorption of metallic adatoms, although it might both gain and loss charge in the interaction with non-/semi-metallic elements depending on the particle size, number of valence electrons and electronegativity.

While, bare $C_3N$ behaves as a semiconductor, the functionalization with adatoms may alter this electronic character into the metallic or half-metallic one. Our DFT results suggest that many metallic elements introduce free electron (n-type doping) to $C_3N$, whereas nonmetallic and metalloidic species mostly create free holes (p-type doping). TM elements embedding *d* electronic states can induce magnetic properties to the monolayer.

Our calculations suggest that functionalized $C_3N$ presents viable electronic-magnetic properties which can be employed in a variety of applications; such as solar cells, photocatalysis, sensors and electronic devices. This study highlights that the chemical functionalization of $C_3N$ is a promising route toward tuning its properties which can consequently motivate further experimental and theoretical studies. With respect to the future theoretical studies, the effect of various parameters on the adatom adsorption such as different DFT functionals and correction schemes; and physical parameters such as adatom-adatom coupling/clustering and the magnetic/electric field effects are highly attractive to be explored.

AUTHOR INFORMATION

**Corresponding Author**

* chandraveer.singh@utoronto.ca

ACKNOWLEDGMENTS

CVS and MM gratefully acknowledge their financial support by University of Toronto, Connaught Global Challenge Award, and Hart Professorship. The computations were carried out through Compute Canada facilities, particularly SciNet and Calcul-Quebec. BM acknowledges support from European Research Council for COMBAT project (Grant number 615132). The authors thank their continued support.

ASSOCIATED CONTENT

Supporting Information

The supporting section includes two tables listing the structural details of $C_3N$ (lattice vectors and positions) and one figure illustrating the total energy as a function of vacuum space.

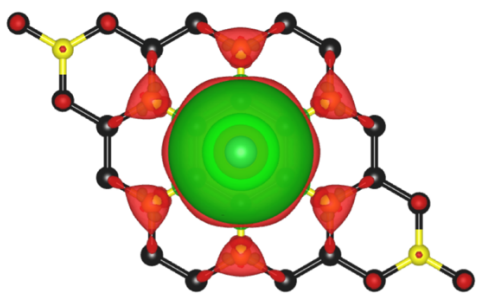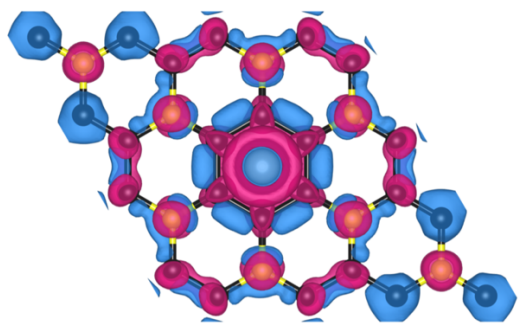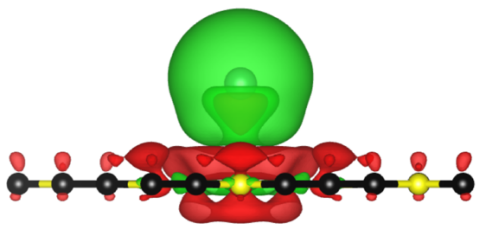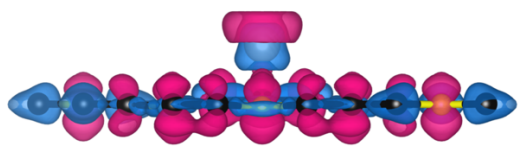

Table of Contents Graphic